\documentstyle[a4,12pt]{article}
 
\renewcommand{\title}[1]{\null\vspace{25mm}
\noindent{\Large{\bf #1}}\vspace{10mm}
}
\newcommand{\authors}[1]{\noindent{\large #1}\vspace{20mm}
    }
\newcommand{\address}[1]{{\center{\noindent #1\vspace{10mm}}
    }}
\renewcommand{\abstract}[1]{\vspace{17mm}
\noindent{\small{\em Abstract.} #1}\vspace{2mm}
   }     
 
\newcommand{\mn}{{\mu\nu}}

\newcommand{\nm}{\nonumber}

\newcommand{\be}{\begin{equation}}
\newcommand{\ee}{\end{equation}}
\newcommand{\ba}{\begin{array}}
\newcommand{\ea}{\end{array}}
\newcommand{\bea}{\begin{eqnarray}}
\newcommand{\eea}{\end{eqnarray}}
\newcommand{\Li}{{\cal L}_\tau}
\newcommand{\id}{i_\tau}
\newcommand{\bh}{\frac{1}{2}}
\newcommand{\bzd}{\frac{2}{3}}

\newcommand{\tr}{{\rm tr}}
\newcommand{\del}{\delta_\tau}
\newcommand{\intM}[1]{\int_{{\cal M}_#1}}
\newcommand{\vide}{\emptyset}

\newcounter{saveeqn}

\begin{document}   \setcounter{table}{0}
 
\begin{titlepage}
\begin{center}
\hspace*{\fill}{{\normalsize \begin{tabular}{r}
{\sf hep-th/0002167}\\
{\sf REF. TUW 99-18}\\
 {\sf LYCEN 2000-03}\\
			      {\sf \today}\\
                              \end{tabular}   }}

\title{Vector supersymmetry in topological field theories.}

\authors {F.~Gieres$^1$, J.~Grimstrup$^2$, 
T.~Pisar$^3$ and M.~Schweda$^4$}    \vspace{-20mm}
       
\address{Institut f\"ur Theoretische Physik,
Technische Universit\"at Wien\\
      Wiedner Hauptstra\ss e 8-10, A-1040 Wien, Austria}
\footnotetext[1]{On sabbatical leave from Institut 
de Physique Nucl\'eaire de Lyon,  
Universit\'e Claude Bernard, 43, boulevard du 11 novembre 1918, 
F-69622-Villeurbanne.}
\footnotetext[2]{Work supported by ``The Danish Research Academy''.}  
\footnotetext[3]{Work supported by the ``Fonds zur F\"orderung der 
Wissenschaflicher Forschung'', 
under Project Grant Number P11582-PHY.}
\footnotetext[4]{email: schweda@hep.itp.tuwien.ac.at}       
\end{center} 
\thispagestyle{empty}

\abstract{We present a simple derivation of
vector supersymmetry transformations 
for topological 
field theories of Schwarz- and Witten-type. 
Our method is similar to the derivation of
BRST-transformations from the 
so-called horizontality conditions
or Russian formulae. 
We show that this procedure reproduces in a concise way 
the known vector supersymmetry transformations
of various topological models and we use it to obtain 
some new transformations of this type for 
$4d$ topological YM-theories 
in different gauges.}

\end{titlepage}



\newpage
\setcounter{page}{2}

\section {Introduction}

It is well known that there exist two classes of 
topological quantum field theories (TQFT's) \cite{brt}.
The so-called {\em Witten-type models}, e.g.  
topological Yang-Mills (YM) theory 
on a four-dimensional space-time manifold 
\cite{Witten:1988ze}-\cite{Brandhuber:1994uf} 
and its higher-dimensional 
generalizations \cite{Baulieu:1998xd} 
and the {\em Schwarz-type models}, e.g. 
the Chern-Simons theory in any odd space-time dimension and 
the BF model in arbitrary dimensions. 
These models are not only invariant 
under BRST-transformations, but also (at least in certain gauges) 
under the so-called 
{\em vector supersymmetry} (VSUSY). The 
anticommutator of 
this transformation
with the BRST-operator yields  
space-time translations (either off-shell or on-shell)  
and thus generates a 
superalgebra of Wess-Zumino type 
\cite{Delduc:1989ft, PiguetSorella, Schweda}.
The VSUSY plays a central role for proving  
the perturbative finiteness of Schwarz-type models
(e.g. see \cite{PiguetSorella}) and it is most 
helpful for discussing
the algebraic renormalization of Witten-type models 
\cite{Brandhuber:1994uf}. 

In reference \cite{PiguetSorella}, 
a general procedure for obtaining 
the explicit form of VSUSY-transforma\-tions 
 was presented 
in the case of three-dimensional Chern-Simons 
theory. However, 
this derivation becomes 
rather involved for more complex models.
Another approach to VSUSY consists of writing the most 
general Ansatz for these transformations  
which is compatible with dimensional and 
covariance constraints and 
subsequently eliminating terms 
by imposing 
the aforementioned anticommutation relations and 
the invariance of the action. 
In practice, this also turns out to be a laborious 
undertaking. 

The aim of the present work is to give 
a short derivation of the two different forms 
of VSUSY's appearing  
in topological models. 
Our results also allow us to cast these transformations
into a compact form. 
The proposed  procedure closely parallels the 
derivation of BRST-transformations in field theories
with local symmetries from the 
so-called horizontality conditions
or Russian formulae
\cite{Baulieu:1984ih, fg,  Bertlmann}. 
The latter enclose all field variations 
in a single (or a few) simple equation(s). 
Though the horizontality conditions 
admit a clear geometric interpretation
in the case of ordinary
YM-theories \cite{Bertlmann}, 
they seem  to be a bit mysterious for 
more general field theoretic models. 
Therefore, we will first provide some insight into their 
working mechanism. In particular, we will 
emphasize that they not only 
encode all information
concerning the explicit form of BRST-transformations, 
but also about their nilpotency.

Our paper is organized as follows. Section 2 is devoted 
to reviewing the horizontality conditions for some of 
the prototype 
models of TQFT mentioned above.
In section 3, we discuss the VSUSY 
before presenting our general
derivation of VSUSY-transformations in section 4. 
Some comments are gathered in a concluding section.
We note that all of our considerations concern the 
classical theory (tree-level).

\section{BRST-symmetry in topological field theories}

As prototype models, we consider 
topological YM-theory on a Riemannian $4$-manifold ${\cal M}_4$, 
Chern-Simons theory on ${\bf R}^3$ and the BF model
on a Riemannian
manifold ${\cal M}_{p+2}$ of dimension $p+2$.
All fields in these models are given by 
differential forms with values in a 
Lie algebra. 
The local symmetries will be described in the BRST-framework, 
i.e. 
infinitesimal symmetry parameters are turned into ghost fields
and symmetry transformations are collected in a 
BRST-transformation. 
Thus, all fields are characterized by their form degree and 
ghost-number which we specify by means of lower and upper indices
for the fields.
All models under consideration involve a  
YM-connection one-form\footnote{Here, 
the $T_a$ represent a basis of the 
Lie algebra and they are assumed to satisfy  
${[ T_a , T_b ]} = {\rm i} f_{ab}^c T_c$ and
$\tr(T_aT_b)=\delta_{ab}$.} $A = A_{\mu}^a T_a dx^{\mu}$. 
The associated field strength is defined by
\be
	F =dA+\bh[A,A],
\label{F2}
\ee
where $[.,.]$ denote graded brackets.
From the nilpotency 
of the exterior derivative $d$, it follows that $F$ satisfies the 
Bianchi identity: 
$D F \equiv 
d F + [A, F ] = 0$. 

It is convenient  to combine the gauge field and the associated 
ghost $c$ as well as the exterior derivative and the BRST-operator
$s$
by  introducing the expressions 
\bea
\label{gde}
	\tilde A &=& A+c
\\
	\tilde d &=& d+s . 
\nm
\eea 
Accordingly, one defines 
\bea
\tilde F &=& \tilde d \tilde A+\bh[\tilde A,\tilde A]
\label{hcW}
\\
\tilde D &=& \tilde d + [\tilde A , \cdot ] .
\nm 
\eea
By definition, the $s$-operator anticommutes with $d$,
and $d$ is nilpotent, henceforth
\bea
\label{nil}
\tilde d^2 =s^2 \qquad \mbox{on all fields}.
\eea

By expanding $\tilde F$ with respect to the ghost-number, 
we find that it has an expression of the form 
\bea
\tilde F = F^0_2 + F^1_1 + F^2_0 , 
\label{expa}
\eea
where 
\bea
F^0_2 &\equiv & F 
\nm \\
\label{ser}
F^1_1 &\equiv & sA + Dc 
\\
F^2_0 &\equiv& sc + \bh [c,c]
.
\nonumber 
\eea 
From equations  (\ref{gde})-(\ref{nil}),
it follows that 
\bea
\tilde D \tilde F = \tilde d^2 \tilde A = s^2 A +s^2 c
\label{co1}
\eea
and similarly  equations (\ref{hcW})-(\ref{expa}) imply
\bea
\tilde D (\tilde D \tilde F) = \tilde d ^2 \tilde F 
= s^2 F +s^2 F^1_1 + s^2 F^2_0 .
\label{co2}
\eea

So far, we have simply derived some equations involving 
$s$-variations of $A$ and $c$ without 
specifying the latter.  
According to equations (\ref{expa}),(\ref{ser}), 
these can be determined by imposing 
a {\em horizontality condition}, i.e. by prescribing 
$F_1^1$ and  $F_0^2$ in equation (\ref{expa}),
$F_2^0$ being necessarily equal to $F \equiv dA + 
\bh [A,A]$. 
Equations (\ref{co1}) and 
(\ref{co2}) then allow us to discuss the nilpotency of the 
resulting BRST-transformations and thereby to check 
the consistency
of the imposed horizontality condition.

\subsection{Witten-type models}

\subsubsection*{Topological Yang-Mills theory}

The classical action reads
\be
	\Sigma^{W}_{inv}=\int_{{\cal M}_4} \tr\left(F F \right),
\label{invW}
\ee
where the wedge product symbol  has been omitted.
 
Due to the {\em shift-} (or topological $Q$-) {\em symmetry}
$\delta A = \psi_1^1$ which  is present in 
this type of model
\cite{Witten:1988ze}-\cite{Brandhuber:1994uf},
the connection $A$ is associated with ghost fields $\psi_1^1$ and 
$\varphi^2_0 \equiv \varphi^2$. 
Henceforth, one imposes the horizontality condition 
\cite{Baulieu:1988xs}
\bea
	\tilde F = F +\psi_1^1+\varphi^2.
\label{hori}
\eea
Substitution of equations (\ref{expa}),(\ref{ser}) 
into this relation
yields the BRST-transformations
\bea
	sA&=&\psi_1^1-Dc \nm \\
	sc&=&\varphi^2- \bh [c,c] .
\label{BRSTW}
\eea

Since $sA$ involves an inhomogeneous term $\psi_1^1$, 
the requirement $s^2 A =0$ determines $s\psi_1^1$
in terms of the other fields 
and analogously the condition $s^2 c =0$ determines $s\varphi^2$.
In order to 
 obtain the explicit form of these variations, 
as well as the one of $F$, 
we note that 
substitution of $s^2 A = 0 = s^2 c$ in  equation (\ref{co1})
yields the generalized Bianchi identity $\tilde D \tilde F=0$.
By expanding the latter with respect to the ghost-number, 
one readily obtains 
\bea 
\label{sf}
	sF  &=& -D\psi_1^1-[c,F ]  \\
	s\psi_1^1 &=&-D\varphi^2-[c,\psi_1^1] \nm \\
	s\varphi^2&=&-[c,\varphi^2]. \nm 
\eea
Finally, substitution of $\tilde D \tilde F=0$
into  equation (\ref{co2}) allows us to conclude  
that $s^2 (F, \psi_1^1, \varphi^2)=0$.
Last, but not least, it can be verified explicitly that 
the  transformation (\ref{sf}) of $F$ leaves  the 
classical action (\ref{invW}) invariant.

\subsection{Schwarz-type models}

\subsubsection*{Chern-Simons theory}

The classical action of this model is given by 
\be
	\Sigma^{CS}_{inv}
=\bh \int_{{\bf R}^3}\tr \, (AdA+\frac{1}{3}A[A,A] ).
\label{invCS}
\ee
Due to the absence of the shift-symmetry
in this model, one imposes
the  horizontality condition  
\be
	\tilde F = F ,
\label{hcCS}
\ee
which implies 
\bea
	sA&=&-Dc \nm \\
	sc&=&-\bh [c,c] .
\label{BRSTCS}
\eea
From equations (\ref{BRSTW}) and (\ref{sf}), 
we see that the  truncation (\ref{hcCS}) of 
(\ref{expa}) is consistent and that it leads to 
nilpotent $s$-variations. 

\subsubsection*{BF model}

Apart from a YM $1$-form $A$ and its associated ghost $c$,
this model involves a $p$-form $B\equiv B_p^0$ 
transforming with the adjoint representation 
of the gauge group.  
Its  field strength $H \equiv DB$ automatically satisfies
the second Bianchi identity $DH= [F,B]$.
The model is characterized by the classical action
\be
	\Sigma^{BF}_{inv}
= \int_{{\cal M}_{p+2}}\tr\left(B F \right),
\label{invBF}
\ee
which is not only invariant under ordinary gauge transformations,
but also under the reducible local symmetry
$\delta B = DB_{p-1}^1$. Henceforth, 
the field $B$ is associated with a series of ghosts 
$B_{p-1}^1, B_{p-2}^2,...,B_{0}^p$
which can be combined in a generalized field 
$\tilde B$, by analogy to the definition of $\tilde A$
in the YM-sector:
\be
	\tilde B = B +B_{p-1}^1+\ldots+B_1^{p-1}+B^p_0 .
\ee
Then, the  generalized field strength 
\[
\tilde H \equiv \tilde D \tilde B
\]
admits the expansion 
\be
\tilde H = H_{p+1}^0 + H_{p}^1 + ... + H^{p+1}_0 ,  
\ee
where
\bea
H_{p+1}^0 & = & DB_p^0
\nm \\
H_{p}^1 & = & D B_{p-1}^1 + sB_p^0 + [c, B_p^0 ]
\label{exh}
\\
& \vdots &
\nm \\
H^{p+1}_0 & = &  sB_0^p + [c, B_0^p ] .
\nm 
\eea

Due to the absence of shift-symmetries, we proceed 
by analogy with the Chern-Simons model and truncate
the expansions of the field strengths, i.e. we impose the 
horizontality conditions
\bea
	\tilde F &=& F \nm \\
\tilde H &= & H.
\label{hbf}
\eea
From equations (\ref{exh}), we then obtain the   
BRST-transformations 
\bea
	sB_{p-q}^q&=& -DB_{p-q-1}^{q+1}-\left[c,B_{p-q}^q \right] 
\hspace{1cm} {\rm for} \ \; q = 0,1,...,  p-1
\nm \\
	sB^p_0 &=&-\left[c,B^p_0 \right],
\label{BRSTBF}
\eea
the transformations of the connection $A$ 
and ghost field $c$ being given by equations (\ref{BRSTCS}).  

In order to check the nilpotency of the $s$-variations  
(\ref{BRSTBF}), we note that 
\[
s^2B +s^2 B_{p-1}^1+ ... + s^2 B^p_0 
\, = \, \tilde d ^2 \tilde B \, = \,
\tilde D \tilde H - [\tilde F ,  \tilde B ] , 
\]
which implies (by matching the ghost-numbers on the 
left and right hand sides) 
\bea
\label{nilos}
	s^2 B^q_{p-q}
& =& -\left[F ,B^{q+2}_{p-q-2}\right] 
\hspace{1cm} {\rm for} \ \, q = 0,1,...,  p-2
\\
	s^2 B^q_{p-q} 
& =& 0
\ \ \, \hspace{3cm}  {\rm for} \ \, q = p-1,  p.
\nm 
\eea
Here, the right hand side vanishes, if we use  
the equation of motion $F=0$ following from the 
classical action (\ref{invBF}). Thus, the $s$-variations 
(\ref{BRSTBF}) are only nilpotent on-shell 
\cite{wallet, PiguetSorella}.
The origin of this result can be drawn back to the fact that
the used horizontality conditions (which were 
motivated by the absence of shift-symmetries)
enforced a truncation of the ghost-expansion  of $\tilde H$,
which is not consistent in the sense that it leads 
to an on-shell algebra.
This kind of phenomenon is familiar from supersymmetric 
field theories where the elimination of auxiliary fields 
from a superfield expansion leads to a supersymmetry algebra 
which only closes on-shell. 
In the BRST-framework, the on-shell closure of the symmetry
algebra is reflected by the fact 
that the $s$-variations of the ghost fields are only nilpotent 
on-shell. 
From our discussion, it followed that this information
 can be directly extracted from the horizontality 
conditions.

\section{Vector supersymmetry}

In flat space-time, infinitesimal VSUSY-transformations are 
parametrized by a constant vector field. 
On curved space-time manifolds, one has to consider 
a covariantly constant vector field 
\cite{feich}. Although this can be done at the expense 
of technical complications, we will limit our 
discussion of VSUSY to flat space-time for the sake of 
simplicity. 

The total action 
$\Sigma = \Sigma_{inv} + \Sigma_{gf}$ 
of a  topological model not only involves classical 
and ghost fields,
but also anti-ghost and Lagrange multiplier
fields\footnote{Here, ``classical'' fields
are not opposed to quantum fields, but simply  
refer to the fields appearing in the classical action.}. 
Let us denote all these fields collectively by
$(\Phi_i)_{i=1,2,...}$. 
Their infinitesimal VSUSY-variations 
$\delta_\tau \Phi_i \equiv \tau^\mu \delta_\mu \Phi_i$
are parametrized by a constant, 
$s$-invariant vector field 
$\tau = \tau^\mu \partial _{\mu}$ of ghost-number zero.
The operator $\delta_{\tau}$ acts as an antiderivation which 
lowers the ghost-number by one unit 
and which anticommutes with $d$.   

The existence and explicit form 
of VSUSY-transformations for a topological model 
described by the action  $\Sigma$ depends, in general,  
in a sensitive way on the choice 
of the gauge-fixing condition. In order to study the existence
of this symmetry and to determine 
the explicit form of VSUSY-transformations, one can  
apply a general procedure 
presented in reference \cite{PiguetSorella}.
This method is based on the 
facts that the gauge-fixing term  
is BRST-exact and that it is 
metric dependent  (while the classical action is 
metric independent): thus, 
the energy-momentum  
tensor is a 
BRST-exact expression,  
\be
	T_\mn=s\Lambda_\mn ,
\ee
which is a typical feature of topological 
models \cite{Witten:1988ze, brt}.
After determining $\Lambda_\mn$,
one expresses $\partial ^{\mu} \Lambda_\mn$
in terms of the functional 
derivatives $\delta \Sigma /\delta \Phi_i$, thereby  
producing 
{\em contact terms}, 
i.e. expressions which vanish when the equations of 
motion are used. 
If 
\be
\partial ^{\mu} \Lambda_\mn \, = \, \mbox{contact terms}
\, + \, \partial ^{\mu} \varepsilon _\mn
, \qquad 
{\rm where} \ \  
s(\partial ^{\mu} \varepsilon_\mn)  =  0,
\ee
then the quantities 
\begin{eqnarray}
\hat \Lambda _\mn &=& \Lambda _\mn - \varepsilon _\mn
\\
\nonumber
\hat T _{\mu\nu} &=& T _{\mu\nu} - s\varepsilon _\mn
\end{eqnarray}
are conserved  
up to equations of motion. They are still related by
$\hat T _{\mu\nu}= s \hat \Lambda _\mn $  
and $\hat T _{\mu\nu}$ can be viewed as an
improvement of the  energy-momentum tensor.
More explicitly, one finds 
\[
\partial^{\nu} \hat \Lambda _{\nu\mu} =
\tr \, ( V_{i\mu} 
\frac{\delta \Sigma}{\delta \Phi_i} \, ) ,
\]
where $V_{i\mu}$ are polynomials in the fields $\Phi _i$
and their derivatives. 
Integration of the last equation over 
the $n$-dimensional space-time on which the topological model
is defined, yields 
\be
0 \, = \, \intM{n} d^nx\; 
\tr \, ( V_{i\mu} 
\frac{\delta \Sigma}{\delta \Phi_i} \, ) .
\label{EnergyCS}
\ee
This relation  expresses the invariance of $\Sigma$ under the 
VSUSY-transformations 
 $\delta_{\mu} \Phi_i := V_{i\mu}$.
A nice feature of this approach is that, by construction, 
the obtained variations of the fields 
represent a symmetry 
of the theory. Yet, for a given TQFT, it  
may be quite tedious to carry out the calculations. 

For all known models, the 
VSUSY- and
BRST-operators satisfy the anticommutation 
relations
\be
	[s,\del]\Phi _i
=\Li\Phi _i \; + \; \mbox{contact terms}. 
\label{algebra}
\ee
Here, $\Li=[\id,d]$ represents 
the Lie derivative along the 
vector field $\tau$ and $\id$ denotes 
the interior product with $\tau$.
Since the algebra closes
on space-time translations, it  
describes a superalgebra of Wess-Zumino type and, 
for brevity, we will refer to (\ref{algebra})
 as the {\em SUSY-algebra}.
More precisely, this algebra closes off-shell 
for Witten-type models and on-shell 
for Schwarz-type models. The lack of off-shell closure
for the latter theories can be explained by the fact that 
the shift-symmetry 
and thereby the associated ghosts are ``missing''. 

In the next section, we will use differential forms to 
present the known, 
as well as some new, results
for our prototype models.  
Before doing so, we summarize some useful algebraic relations.  

The graded commutation relations between the basic operators
read:
\bea
0  & =& {[ s , d ]}  \; =\;
{[ s , i_{\tau} ]}  \; =\; {[ s , {\cal L} _{\tau} ]}
\label{grac}
\\
0 & =& {[ \delta _{\tau}, d ]} 
\; =\; {[ \delta _{\tau}, \id  ]} 
\; =\;
{[ \delta _{\tau}, \Li ]} .
\nm 
\eea

As usual, the Hodge dual of a Lie algebra-valued 
differential form $\Omega$ will be 
denoted by $*\Omega$. 
On a $n$-dimensional space-time manifold 
${\cal M}_n$, the star operator can be used to define 
a scalar product of Lie algebra-valued 
$p$-forms $\Omega_p^q$ which, in addition,
have some ghost-number $q$: 
\[
\langle \Omega_p^q , \Lambda_p^r \rangle \equiv 
\int_{{\cal M}_n}
{\rm tr} \, ( \Omega_p^q \, * \! \Lambda_p^r ).
\]
This product has the graded symmetry
\bea
\langle	\Omega_p^q, \Lambda_p^r \rangle
=(-1)^{(p+n)(q+r)+qr} \langle \Lambda_p^r , \Omega_p^q \rangle .
\eea
If the space-time dimension is odd, the star operation 
represents a mapping between forms of even and odd degree, 
henceforth it anticommutes 
with the antiderivation $s$. 

By using the metric tensor $(g_{\mn})$ of the 
space-time manifold, one can associate
the $1$-form 
$g(\tau) \equiv \tau ^{\mu} g_{\mn} dx^{\nu}$ to the vector field 
$\tau = \tau^{\mu} \partial_{\mu}$.   
(In the 
mathematical literature, this mapping and its 
inverse are known as the ``musical isomorphisms'', which are 
usually denoted by $\flat$ and $\sharp$, 
respectively \cite{bmoll}.)  
The Hodge operator intertwines between the 
interior product $\id$ and the exterior multiplication 
with $g(\tau)$:
\bea
* \; g(\tau) \, \Omega_p^q &=& (-1)^p \ \id \! *  \Omega_p^q 
\label{inter}
\\
g(\tau) * \, \Omega_p^q &=& (-1)^{p+1} \, 
* \id \Omega_p^q. 
\nm
\eea

\subsection{Witten-type models}

In the literature, two different types of gauge-fixings 
have been considered for topological YM-theory.  
\begin{itemize}
\item{The first choice consists of a linear gauge condition
for both the shift-symmetry and the ordinary gauge symmetry 
\cite{Brandhuber:1994uf}:
\be
\label{bra1}
\Sigma^{W} _1 \, = \, 
\int_{{\cal M}_4} \tr\left(F F \right)
\, + \, s \int_{{\cal M}_4} \tr\left\{
\chi_2^{-1} F^+ + \bar{\phi}^{-2} d * \! \psi_1^1
+ \bar c d * \! \! A \right\} .
\ee
Here, the fields $F^+ \equiv {1\over 2} (F + *F),   
\, \chi_2^{-1}$ and $s\chi_2^{-1} \equiv B_2$ 
are self-dual and the BRST-variations
are defined by (\ref{BRSTW}),(\ref{sf}) and 
\be
\ba{rclcrcl}
s \chi_2^{-1} &=& B_2
&,&
sB_2 &=& 0 
\nm \\
s\bar{\phi}^{-2}&=& \eta^{-1} 
&,&
s\eta^{-1} &=&0
\label{saux}
\\
s\bar c &=& b 
&,&
sb&=&0 .
\nm
\ea
\ee
The action (\ref{bra1}) is also invariant under the following 
VSUSY-variations 
\cite{Brandhuber:1994uf}:
\bea
	\del A &=& 0 
\quad , \quad 
\del \psi_1^1 \; =\; \Li A 
\nm \\
	\del c &=& 0 
\quad , \quad 
	\del \varphi^2 \; = \; \Li c
\label{susyW1}
\eea
and 
\be
\ba{rclcrcl}
\del\bar c &=&-\Li\bar\phi^{-2}
& , & \del b&=&\Li\bar c + \Li\eta^{-1}
\nm\\
	\del\chi_2^{-1}&=&0
&,&\del B_2&=&\Li\chi_2^{-1} 
\nm \\
	\del\bar\phi^{-2}&=&0
&, & \del \eta^{-1}&=&\Li\bar\phi^{-2}.
\label{susy2}
\ea
\ee }
\item{The second choice is as follows 
\cite{Baulieu:1988xs,Myers:1990ur,Ouvry:1989mm}. 
For the shift-symmetry, 
one considers 
a covariant gauge condition and  for the ordinary gauge symmetry, 
one chooses either 
{\em (a)} a linear, 
{\em (b)}  a covariant or 
{\em (c)}  no gauge condition at all:
\be
\label{bra2}
\Sigma^{W} _{2\alpha} \, = \, 
\int_{{\cal M}_4} \tr\left(F F \right)
\, + \, s \int_{{\cal M}_4} \tr\left\{
\chi_2^{-1} F^+ + \bar{\phi}^{-2} D* \! \psi_1^1 \right\}
+s\Psi _{\alpha} 
\quad {\rm with}
\ \, \alpha \in \{ a,b,c \},
\ee
where 
\bea
\Psi_a=\int_{{\cal M}_4} \tr\left\{\bar cd*\!\!A \right\}
\quad , \quad 
\Psi _b=\int_{{\cal M}_4} \tr\left\{\bar cD*\!\!A \right\}
\quad , \quad 
\Psi _c=0.
\eea
The BRST-transformations are again given by 
(\ref{BRSTW}),(\ref{sf}) and (\ref{saux}) (or by adding a 
gauge symmetry contribution $-[c, \Phi_i]$ to each variation
$s\Phi_i$ in (\ref{saux}) - such a contribution does not matter for 
our considerations). 

For the action (\ref{bra2}), we have found 
the VSUSY-variations
\bea
	\del A &=& 0 
\quad \quad , \quad 
\del \psi_1^1 \; =\; \id F =  \id d A + [\id A ,A]
\nm \\
	\del c &=& \id A
\quad , \quad 
	\del \varphi^2 \; = \; \id \psi_1^1
\label{susyW2}
\eea
and 
\be
\ba{rclcrcl}
\del\bar c&=&0
&, & \del b&=&\Li\bar c
\nm \\
\del\chi_2^{-1}&=& 2 \left( \id*\!D\bar\phi^{-2} \right) ^+
&,& \del B_2&=&\Li\chi_{2}^{-1}
+2 \left( \id*\! ( [\psi_1^1-Dc,\bar\phi^{-2}] 
- D\eta^{-1}) \right) ^+
\nm \\
\del\bar\phi^{-2}&=&0
&,& \del \eta^{-1}&=&\Li\bar\phi^{-2}.
\label{susy1}
\ea
\ee
By contrast to the transformations (\ref{susyW1}),(\ref{susy2}), 
these
variations are {\em not} 
linear in the basic fields. 
Since $A$ and $\bar c$ do not transform under $\del$, 
the gauge-fixing term $s\Psi _{\alpha}$ for ordinary gauge 
symmetry is, taken by itself, invariant under VSUSY.
In section 4.1 below, we will explain how we determined the given 
VSUSY-transformations and why 
they represent a symmetry of the action.}  
\end{itemize}

Both sets of 
VSUSY-transformations have several features in common,
which can thereby be considered as characteristic for 
topological models of Witten-type. 
First,  both of them fulfill the SUSY-algebra (\ref{algebra}) 
off-shell. 
Second,  both of them leave the classical field $A$ inert
 and therefore they 
do not act on the classical 
action. Thus, they
only represent
a non-trivial symmetry of the gauge-fixing part of the action. 
This should explain the fact that the VSUSY is not restrictive 
enough for topological YM-theory
to make the model perturbatively finite, though 
its existence considerably improves the
algebraic renormalization procedure, 
leading to an anomaly-free quantized theory
\cite{Brandhuber:1994uf}.
Finally, we remark that the 
classical and ghost fields, i.e. the fields 
which belong to the geometric part of the BRST-algebra,
transform among themselves under VSUSY:
none of the anti-ghosts or Lagrange multipliers 
involved in the gauge-fixing action appears
in the variations (\ref{susyW1}) and (\ref{susyW2}).

\subsection{Schwarz-type models}

\subsubsection*{Chern-Simons theory}

	In the Landau gauge, 
the total Chern-Simons action is given by 
\cite{Delduc:1989ft,PiguetSorella}
\bea
	\Sigma^{CS}&=&\Sigma_{inv}^{CS}
\, + \, \Sigma_{gf}^{CS} \nm \\
	&=&\bh \int_{{\bf R}^3} \tr \, (AdA+\bzd A^3 )
+\int_{{\bf R}^3} \tr \left\{ bd *\!\!A
+\bar cd*\!\! Dc \right\} ,
\label{CSaction}
\eea
where $\bar c$ and $b$ are, 
respectively, the anti-ghost and the corresponding 
Lagrange multiplier, both forming a BRST-doublet:
$s \bar c =b, \, sb=0$.

Substitution of  
the functional derivatives of $\Sigma^{CS}$ with respect
to $A, b$ and $c$, e.g.    
\bea
\frac{\delta \Sigma^{CS}}{\delta A}&=&F -*db-[c,*d\bar c], 
\label{eomCS}
\eea
into expression (\ref{EnergyCS}), 
leads to the (linear)
VSUSY-variations \cite{Delduc:1989ft,PiguetSorella}
\bea
	\del A 
&=& -\id *\!d\bar c  
\quad \, , \quad 
	\del \bar c \; = \; 0 
\nm \\
	\del c&=&\id A 
\qquad \quad , \quad 
	\del b \; =\; \Li \bar c .
\label{kr}
\eea
The SUSY-algebra (\ref{algebra}) now closes off-shell 
for $c,\bar c$ and $b$,
but not for  the classical field $A$:
\be
\label{known}
	[s,\del]A=\Li A
-\id \frac{\delta \Sigma^{CS}}{\delta A}.
\ee

From these results (and similar results 
for the BF model discussed below), 
we conclude that the VSUSY-transformations 
in Schwarz-type models differ substantially from those 
in Witten-type models: the classical fields are not invariant,
but transform into anti-ghost fields, and 
the SUSY-algebra does not close 
off-shell for the classical fields.
The fact that 
 the transformations
mix the classical and gauge-fixing parts of the total action
renders the VSUSY highly non-trivial 
and constraining for the quantum theory: 
it is at the origin of the 
perturbative finiteness of these models. 

We note that a (linear) VSUSY is also present if the axial gauge 
is chosen \cite{Brandhuber:1993ur, Schweda}, 
but it does not exist for a covariant gauge 
\[
{\delta \Sigma \over \delta b} = d *\!\!A + \alpha b 
\qquad (\alpha \in {\bf R}^*).
\]
The fact that the presence of VSUSY implies a certain 
class of gauges 
is a feature that is reminiscent of the anti-BRST symmetry, 
whose presence has similar consequences (if considered 
in addition to the usual BRST-symmetry) \cite{fg}. 
However, the VSUSY has a considerably
richer structure which entails interesting results for 
the quantum theory, which is not the case for the anti-BRST 
symmetry.

\subsubsection*{BF model}
 
The total BF action 
$\Sigma^{BF}=\Sigma^{BF}_{inv}+\Sigma^{BF}_{gf}$ involves 
\[
	\Sigma^{BF}_{gf}= s \int _{{\cal M}_{p+2}} 
\tr\left\{ \bar cd*\!\!A+
\bar c_{p-1}d*\!\!B +\ldots \right\}, 
\]
where $\bar c$ and 
$\bar c_{p-1}$ are the anti-ghosts which fix the Landau gauge 
in the YM- and $B$-field sector, respectively,
and where we only 
wrote out the terms which are relevant here.
The derivation of VSUSY-transformations proceeds along the lines 
of the Chern-Simons theory, 
though one has to take into account the fact that the 
$s$-variations (\ref{BRSTBF}) of the BF model 
are only nilpotent by virtue of the classical 
equation of motion $F=0$ (cf. equations (\ref{nilos})): 
since we are now dealing with 
the complete, gauge-fixed action, these 
$s$-variations have to be extended 
in an appropriate way so as to relate to the 
complete equation of motion.
This can be achieved by 
standard methods \cite{wallet, PiguetSorella}  
and the following VSUSY-transformations can be 
found \cite{PiguetSorella}: 
\bea
	\del A &=&-\id *\! d\bar c_{p-1}
\nm \\
	\del B &=& (-1)^p\id *\! d\bar c
\label{susyBF}\\
\del B_{p-k}^k &=& \id B^{k-1}_{p-k+1}
\qquad \quad {\rm for} \ \; k= 1,..., p . 
\nm
\eea
For the classical fields, the SUSY-algebra (\ref{algebra})
only closes on-shell:
\bea
	\left[s,\del\right]A&=&\Li A-\id 
\frac{\delta \Sigma^{BF}}{\delta B} \nm \\
	\left[s,\del\right]B
&=&\Li B-\id \frac{\delta \Sigma^{BF}}{\delta A}.
\label{algebraBF}
\eea

\section{Derivation of vector supersymmetry transformations}

In the sequel, we will repeatedly refer to the quantity
\be
\id \tilde F \, = \, \id (\tilde d \tilde A  
+ \bh [\tilde A , \tilde A ])
\, = \, \id \tilde d \tilde A - [ \tilde A ,\id \tilde A ] .
\nm 
\ee
By virtue of 
$[\id, \tilde d]  = [\id, d+s] = \Li$ and $\tilde D = \tilde d
+ [\tilde A,\cdot \, ]$, this expression takes the compact form   
\be
	\id\tilde F
=\Li\tilde A-\tilde D\id\tilde A . 
\label{if}
\ee

We now present an alternative approach 
to the derivation of  VSUSY-transformations for 
topological models. 
To stress the analogy with the method of 
horizontality conditions for the derivation 
of BRST-transformations, we briefly summarize   
the main steps which are followed for the latter derivation in  
the case of Chern-Simons (or ordinary YM-) theory. 
One {\em assumes} that the BRST-operator 
is nilpotent on the fields $A$ and $c$, i.e.
that the graded algebra 
\be
\label{ss}
[s,s] = 0 
\ee
holds for these fields. 
Then, one {\em postulates} the horizontality conditions 
involving the generalized field strength
$ \tilde F \equiv \tilde d \tilde A + \bh [\tilde A ,\tilde A]$, 
i.e. for Chern-Simons theory,  
one postulates $ \tilde F = F$.
By expanding this relation with respect to the ghost-number,
one immediately obtains the BRST-transformations. 
Their off-shell nilpotency, i.e. the consistency of the 
final equations with the starting point (\ref{ss}), 
 can either be checked explicitly   
or by resorting to the general arguments 
indicated in section 2. As a last step, the invariance 
of a given action is to be verified.
(Eventually, we can also reverse the problem 
and look for action functionals admitting the derived 
BRST-variations as symmetries.)  

Let us now proceed with VSUSY. 
First of all, 
we {\em assume} that the SUSY-algebra (\ref{algebra}) is  
satisfied  {\em off-shell}.  Next, we evaluate the   
$\del$-variation of $\tilde F $:
\bea
	\del \tilde F
&=& \del \tilde d \tilde A - [ \tilde A ,\del \tilde A ]
\nm \\
&=& [ \del, \tilde d]  \tilde A  
- \tilde d \del  \tilde A - [ \tilde A ,\del \tilde A ]
\nm \\
&=& [ \del, \tilde d]  \tilde A  
- \tilde D \del  \tilde A . 
\label{gre}
\eea
Substitution of the assumed off-shell algebra entails
\be
	\del\tilde F
=\Li\tilde A-\tilde D\del\tilde A .
\label{W2}
\ee
Comparison of the expressions (\ref{if}) and (\ref{W2})
now motivates us to 
 {\em postulate} either 
$\vide${\em  -type symmetry conditions},
\bea
	\del \tilde A&=&\id \tilde A
\label{Wsusy2a}
\\
	\del \tilde F &=&\id \tilde F , 
\nm 
\eea
or 
$0${\em -type symmetry conditions}, 
\bea
	\del \tilde A&=&0 \nm \\
	\del \tilde F &=&\Li\tilde A.
\label{Wsusy1}
\eea
In both sets of equations, 
the second relation is a consequence of
(or consistency condition for) the first one 
by virtue of equations (\ref{if}),(\ref{W2}). 
The terminology $\vide$ versus $0$ simply expresses the fact
that $\del \tilde A \neq 0$ as opposed to $\del \tilde A = 0$.

For the $B$-field sector, we follow the same line of reasoning.
From the definition $\tilde H = \tilde D \tilde B$, it 
follows that 
\be
	\id\tilde H
=\Li\tilde B-\tilde D\id\tilde B 
+ [\id \tilde A, \tilde B ],
\label{ib}
\ee
while the assumption that $\tilde B$ satisfies the SUSY-algebra
off-shell, i.e. 
$[s, \del ] \tilde B = \Li \tilde B$, 
leads to 
\be
	\del\tilde H
=\Li\tilde B-\tilde D\del\tilde B
+ [\del \tilde A, \tilde B ].
\label{W4}
\ee
Comparison of both relations then motivates us again to 
{\em postulate} either 
$\vide${\em  -type symmetry conditions},
\bea
	\del \tilde B&=&\id \tilde B
\label{cob}
\\
	\del \tilde H &=&\id \tilde H , 
\nm 
\eea
in conjunction with equations (\ref{Wsusy2a}), or 
$0${\em -type symmetry conditions}, 
\bea
	\del \tilde B&=&0 
\\
	\del \tilde H &=&\Li \tilde B , 
\nm 
\eea
in conjunction with equations (\ref{Wsusy1}).

\subsection{Witten-type models}

Let us substitute $\tilde A = A+ c$ and 
the horizontality condition for topological 
YM-theories, i.e. 
$\tilde F = F + \psi_1^1 + \varphi^2$, into the $0$-type 
symmetry conditions 
(\ref{Wsusy1}). By decomposing  
with respect to the ghost-number, we 
immediately obtain the VSUSY-transformations (\ref{susyW1}).
Similarly, from  
the $\vide$-type symmetry conditions 
(\ref{Wsusy2a}), we reproduce the non-linear VSUSY-transformations 
(\ref{susyW2}). 
(Actually, this is how we found these variations!) 
Thus, the two representations of 
VSUSY defined, respectively, by the $0$-type and 
$\vide$-type symmetry conditions, 
manifest themselves in 
Witten-type models, the symmetry depending 
on the chosen gauge-fixing conditions.  

Let us now come back to the VSUSY-variations (\ref{susy1}) of the 
anti-ghosts and Lagrange multipliers.
The transformations of the  anti-ghosts can be found by assuming 
that the off-shell SUSY-algebra $[s, \del ]= \Li$
is valid, and by varying the gauge-fixing 
action $\Sigma_{gf} \equiv s\int L$:
\[
\del \Sigma_{gf} = \int \del s L = \int \Li L
- s \int \del  L
\] 
By choosing the $\del$-variations 
of the anti-ghosts $( \bar c ,\chi^{-1}_2, \bar{\phi}^{-2})$
in an appropriate way, the last term vanishes and thereby 
ensures the 
$\del$-invariance of $\Sigma_{gf}$. Finally, the 
transformations of the Lagrange multipliers 
$(b, B_2, \eta^{-1})$ are also determined by imposing  
the VSUSY-algebra for them, e.g. from 
\[
\del b= \del (s \bar c) = \Li \bar c - s(\del \bar c)
\]
and the known $\del$-variation of $\bar c$, one finds
the one of $b$.  
Thus, it is by construction that the 
VSUSY-transformations (\ref{susyW2}),(\ref{susy1})
represent a symmetry of the action (\ref{bra2}).
(The same arguments can be used to determine the 
$\del$-variations (\ref{susy2}) and to check the $\del$-invariance 
of the action (\ref{bra1}).)

We also applied our procedure 
to higher-dimensional TQFT's of Witten-type, 
in particular to the six-dimensional 
model of reference \cite{Baulieu:1998xd}. 
In this case as well, we could 
determine the corresponding VSUSY-transformations
in a straightforward way \cite{Ita:1999mx}, 
thereby confirming 
the usefulness of the approach to VSUSY 
outlined here.

\subsection{Schwarz-type models}

Before discussing the examples, we should note right away
that the transformation laws 
that we will derive from the symmetry conditions 
in the present case, are to be considered with caution. 
In fact, our symmetry conditions are based 
on the assumption that the 
SUSY-algebra closes off-shell and this is not the case for 
the classical fields occurring in Schwarz-type models. 

\subsubsection*{Chern-Simons theory}

If we were to combine the horizontality conditions of the 
present model, i.e. $\tilde F = F$, with the $0$-type 
symmetry conditions  (\ref{Wsusy1}), we would obtain 
the inadmissible result $\Li A = 0= \Li c$. Henceforth, the 
$0$-type symmetry conditions (\ref{Wsusy1}) can only occur 
for Witten-type models where a shift-symmetry exists.

Thus, let us apply the $\vide$-type 
symmetry conditions (\ref{Wsusy2a}): by decomposing the 
latter with respect to the ghost-number and by using 
$\tilde F = F$, we get 
\be
\label{inc}
\del A =0 
\qquad , \qquad 
\del c = \id A  	
\ee
and 
\be 
\label{buh}
\del F  = 0 
\qquad , \qquad   
\id F =0 . 
\ee
From the transformations (\ref{inc}) 
and the $s$-variations of $A$ and $c$, 
it follows that 
\bea
{[s , \del ]} A & = & \Li A - \id F  
\label{alclas} 
\\
{[s , \del ]}c & = & \Li c ,
\nm
\eea
where $\id F=0$, if the classical 
equations of motion are used.
Thus, we have obtained an on-shell algebra after having assumed
the validity of an off-shell algebra as our starting point:
this result is due to the truncation of the ghost-expansion 
$\tilde F$. 

The algebra (\ref{alclas}) can be interpreted as follows. 
If we only consider the classical action, the latter 
is invariant 
under the $\del$-variations (\ref{inc}) which satisfy 
the SUSY-algebra on-shell.  
We will now try to promote this  trivial symmetry
of the classical action to a non-trivial symmetry 
of the total action (again allowing for an on-shell closure
of the SUSY-algebra). To do so, we retain the 
non-trivial transformation law $\del c = \id A$
and we consider $\del A$ to be unknown. 

Let us evaluate 
the expression $\del sA$ in terms of $\del A$: 
by substituting  the known expressions of  
$sA$ and $\del c$, and 
by using $d\id = \Li -d\id$ as well as $dA + \bh [A,A]=F$, we
obtain 
\be
\label{last}
\del sA = \Li A - \id F + [c, \del A ].
\ee
By virtue of the complete equation of motion for $A$, 
as given by equation (\ref{eomCS}), 
the  classical contact term $\id F$ 
in equation (\ref{last}) can be
expressed in terms of the contact 
term
$\id (\delta \Sigma ^{CS} / \delta A)$. 
(In this way, the anti-ghosts enter 
our geometric framework which only involves classical
and ghost fields.)
Subsequent use of $b =s\bar c$ then  
entails
\[
\del sA = \Li A 
-\id \frac{\delta \Sigma ^{CS}}{\delta A}  
- \id (* ds\bar c) + [c, \id (*  d\bar c)] + [c, \del A ] . 
\] 
By adding the unknown quantity $s\del A$
to both sides of this equation,
we get the result
\be
[s, \del] A = \Li A -\id \frac{\delta \Sigma ^{CS}}{\delta A}
+ s \{ \del A + \id (*  d\bar c)\} 
+ [c, \del A + \id (*  d\bar c)] . 
\ee
Obviously, the choice
\be
\del  A = - \id (*  d\bar c) 
\ee
ensures the validity of the SUSY-algebra (\ref{algebra})
and gives the known results (\ref{kr}),(\ref{known}).
The requirement of invariance of the Chern-Simons
action under the determined $\del$-variations
fixes the transformation laws of $\bar c$ and $b$, 
again in agreement with equations (\ref{kr}). 
(We could also argue that $\del \bar c$
has to vanish for dimensional reasons; then 
$\del b$ follows again by imposing the SUSY-algebra
on $\bar c =sb$.)

Let us summarize once more our procedure:
by starting from the $\vide$-type symmetry conditions, 
we could derive the VSUSY-transformations for 
the Chern-Simons theory solely   
from the knowledge of the total action 
and BRST-transformations and by assuming that the 
SUSY-algebra is fulfilled up to contact terms. 
  
As we have shown in the appendix, the 
$\del$-transformations of $A$ and $c$ can also be 
obtained in a direct way by redoing our initial 
derivation (\ref{gre})-(\ref{Wsusy2a})
after having determined the contact terms 
in the SUSY-algebra by dimensional arguments.

\subsubsection*{BF model}

One proceeds as for the Chern-Simons theory.
If we only consider the classical action, 
the $\vide$-type symmetry conditions (\ref{Wsusy2a}) 
and  (\ref{cob})
lead to equations (\ref{inc}) and to the following variations 
of the $B$-fields:
\bea
	\del B&=&0 \nm \\
	\del B_{p-k}^k&=&\id B_{p-k+1}^{k-1}
\qquad \quad {\rm for} \ \; k= 1,..., p . 
\eea
When extending 
these results to the complete gauge-fixed action, 
one has to take into account the fact that 
the SUSY-algebra is only valid  on-shell and
that it involves   
the complete equations of motion.
By modifying the transformation laws  
$\del A=0$ and $\del B=0$ along the lines indicated above, 
one obtains the VSUSY-transformations (\ref{susyBF}) 
which fulfill the on-shell algebra (\ref{algebraBF}). 

\section{Conclusion }

From the previous considerations, we conclude that 
the VSUSY-transformations for Witten-type models 
follow straightforwardly from 
the $0$-type or $\vide$-type symmetry conditions  
(their presence depending on which gauge-fixing 
condition is chosen). 
This derivation seems to be quite efficient,
in particular for higher-dimensional 
TQFT's \cite{Ita:1999mx}. 

The VSUSY-transformations for Schwarz-type models follow
from $\del\tilde A=\id \tilde A$ 
by checking the algebra and 
by taking into account the equations 
of motion of the model under consideration. 
An off-shell formulation for these theories 
can be obtained by considering the linearized Slavnov-Taylor 
operator which involves external sources.
These sources  are  associated with 
the non-linear terms in the BRST-transformations and they
also transform under the VSUSY which is now  
linearly broken \cite{PiguetSorella}. 
Since the 
Batalin-Vilkovisky formalism naturally incorporates  
sources under the disguise of anti-fields
(e.g. see references 
\cite{Baulieu:1995bq} for 
the application to topological 
models), 
it should represent a more convenient framework for  
discussing Schwarz-type models.
This will be reported upon
elsewhere \cite{wip}.

\vskip 1.2truecm
 
{\bf \Large Acknowledgments}
 
\vspace{3mm}
 
F.G. wishes to thank R.Bertlmann for a stimulating discussion 
on the BF model and he 
expresses his gratitude to all the members 
of the Institut f\"ur Theoretische Physik of the 
Technical University of Vienna
for the warm hospitality extended to him.

\newpage 

\appendix
\section{Another derivation of VSUSY-variations
in Chern-Simons theory} 

In the following, we present a slightly different 
derivation of VSUSY-transformations for Schwarz-type models
by using the Chern-Simons 
theory as an example. This approach is motivated by the 
results (\ref{inc}),(\ref{buh})  
which entail the vanishing of $F$,  i.e. the 
classical, rather than the complete  
equation  of motion for Chern-Simons theory.
This fact exhibits the inadequacy
of our starting point, i.e. of the assumption that the 
SUSY-algebra is fulfilled off-shell. 
Hence, we simply review the derivation 
(\ref{gre})-(\ref{Wsusy2a})
after having determined 
the contact terms 
in the SUSY-algebra (\ref{algebra}) for the commutators 
$[s, \delta_{\tau}] A$ and $[s, \delta_{\tau}] c$. 
These terms can be found 
without explicitly 
knowing the $\del$-variations
since their form is strongly constrained: 
for the commutator
$[s, \delta_{\tau}] A$, this term  has to be 
a function of the functional derivatives
of $\Sigma^{CS}$ with respect to the fields of the model,  
and this function  
must be linear in $\tau$ and of  
the same dimension 
and ghost-number as $A$ (similarly for 
the contact term in the commutator $[s, \del ] c$).  
From these arguments, we can deduce that the algebra can only
have the form 
\bea
	{[ s, \del ]} A 
& = & \Li A 
\, + \, 
\xi \,  \id \frac{\delta \Sigma^{CS}}{\delta A} 
\nm \\
	{[ s, \del ]} c & = & \Li c , 
\nm 
\eea 
where $\xi$ is a real factor. 
We now collect these two equations into a single one 
involving $\tilde A = A + c$ 
and we substitute the known expression (\ref{eomCS}) for the 
functional derivative:
\bea
	[s,\del] \tilde A \, = \, \Li \tilde A \, + \, 
\xi \,  \id \{ F 
+ s (* d\bar c)
-[c, *  d\bar c] \} . 
\eea
This relation represents the correct 
form of the SUSY-algebra for the present model. Thus, 
we substitute it in the $\del$-variation (\ref{gre})
of $\tilde F$:  
\[
\del \tilde F = \Li \tilde A 
+ \xi \,  \id F  
+ \xi \,  \{\id s ( * d\bar c)
-  [c, \id  (* d\bar c)] \}  
- \tilde D ( \del \tilde A).
\]
Next, we substitute the horizontality condition $ \tilde F =F$ 
in $\id F$ 
and eliminate $\Li \tilde A$ by means of 
the general relation (\ref{if}): if $\xi =-1$, 
the $\id F$-term drops out from the 
last equation and we are left with 
\[
\del \tilde F =  \tilde D ( \id \tilde A)
+ s\id  (* d\bar c)
-  [c, \id(*  d\bar c)]   
- \tilde D ( \del \tilde A). 
\]
Thanks to $\tilde D  
\equiv \tilde d + [ \tilde A, \cdot \, ] 
= D + s + [ c, \cdot \, ]$, this result can be rewritten as 
\be 
\del \tilde F 
= - \tilde D \left[  \del \tilde A - \id \tilde A
+ \id(* d\bar c) \right] + D  \id(* d\bar c )  .
\ee
Henceforth, the postulated conditions (\ref{Wsusy1}), 
which did not take into account 
the equations of motion, 
should be modified according to 
\bea
	\del \tilde A&=& \id \tilde A -\id (* d\bar c)
\\
	\del \tilde F&=& D\id(* d\bar c )  .
\nm
\eea
Expansion with respect to the ghost-number
now yields the known results (\ref{kr}).

In summary, the key point of this model-dependent derivation
was to determine the general form 
of contact terms in the SUSY-algebra
for the considered model.


\providecommand{\href}[2]{#2}\begingroup\raggedright\endgroup

\end{document}